\documentclass[fleqn,twosiide]{article}
\usepackage{espcrc2}
\usepackage{graphicx}

\title{Enhanced formation of a dense $\bar{K}$ nuclear cluster $K^- pp$ in $pp$ collisions\\
- $\Lambda^*p$ doorway dominance}

\author{
Toshimitsu Yamazaki\address{Department of Physics, University of Tokyo, Bunkyu-ku, Tokyo 113-0033, Japan, and
RIKEN, Wako, Saitama 351-0198, Japan}
\thanks{E-mail address: yamazaki@nucl.phys.s.u-tokyo.ac.jp} and 
Yoshinori Akaishi\address{College of Science and Technology, Nihon University, Funabashi, Chiba 274-8501, Japan, and
RIKEN, Wako, Saitama 351-0198, Japan}
\thanks{E-mail address: akaishi@post.kek.jp}
}
\begin{document}
\date{\today}


\date{\today}

\begin{abstract}
 We have found theoretically that the elementary process, $p + p \rightarrow K^+ + \Lambda(1405) + p$, 
which occurs in a short impact parameter ($\sim$ 0.2 fm) and with a large momentum transfer ($Q \sim 1.6$ GeV/$c$), leads to unusually large self-trapping of $\Lambda(1405) (\equiv \Lambda^*)$ by the projectile proton, when a $\Lambda^* -p$ system exists as a dense bound state (size $\sim$ 1.0 fm) propagating to $K^-pp$. The seed, called ``$\Lambda^* p$ doorway", is expected to play an important role in the ($p, K^{+,0})$ type reactions and heavy-ion collisions to produce various $\bar{K}$  nuclear clusters.
\end{abstract}


\maketitle

\section{Introduction}

\begin{figure}[htb]
\centering
\includegraphics[width=7cm]{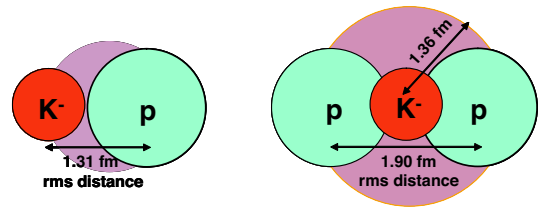}
\vspace{-1.5cm}
\caption{\label{fig:Kp-Kpp} 
 Predicted structure of $K^-p$ and $K^-pp$.
}
\end{figure}

Recently, exotic nuclear systems involving a $\bar{K}$  ($K^-$ and $\bar{K}^0$) as a constituent have been predicted based on phenomenologically constructed $\bar{K}N$ interactions \cite{Akaishi:02,Yamazaki:02,Dote:04a,Dote:04b,Yamazaki:04,Akaishi:05}. The predicted bound states in
$K^-ppn$, $K^-ppnn$ and $K^-$$^8$Be with large binding energies lie below the $\Sigma \pi$ emission threshold, and thus are expected to have narrow decay widths. Because of the strong $\bar{K}N$ attraction they have enormously high nucleon densities, $\rho_{ave} \sim 0.5$ fm$^{-3}$, about 3 times the normal nuclear density $\rho_0  \sim 0.17$ fm$^{-3}$. Such compact nuclear systems, which can be called  ``$\bar{K}$  nuclear clusters" ({\it KNC}), are $\bar{K}$  bound states in non-existing nuclei. The basic ingredient for this new family of nuclear states is $I=0~K^-p$, which is identified as the known $\Lambda (1405)$ (hereafter, expressed as $\Lambda^*$) with a binding energy of $B_K$ = 27 MeV and a width of $\Gamma$ = 40 MeV. The lightest system is $K^-pp$ (and its isospin partner $\bar{K}^0 pn$), which was predicted to have $B_K$ = 48 MeV and $\Gamma$ = 61 MeV  \cite{Yamazaki:02}. This species, which can be called {\it nuclear kaonic hydrogen molecule}, results from a fusion of $\Lambda^*$ and $p$, namely, $\Lambda^*$ as a bound state of $K^- p$ dissolves into a $\bar{K}$  bound state, $K^- pp$,
\begin{equation}
 \Lambda^* p \rightarrow K^-pp. 
\end{equation}
The predicted rms distance of the $\Lambda^*-p$ system is 
\begin{equation}
R(\Lambda^*p) = 1.05~{\rm fm}.
\end{equation}
The structure of $K^-p$ and $K^-pp$ with calculated rms distances of $K^-$-$p$ and $p$-$p$ and rms radius of the $K^-$ distribution is shown in Fig.~\ref{fig:Kp-Kpp}.

Whereas some evidences for kaonic nuclear clusters for $T=1$ $K^-pnn$ ($B_K = 194$ MeV) \cite{Suzuki:04} and $K^-pp$ ($B_K$ = 115 MeV) \cite{FINUDA:PRL} were reported, it is important to produce various $\bar{K}$ clusters by different nuclear reactions and thereby to examine their strucure and properties. A method to determine the sizes of the $\bar{K}$  clusters via the momentum correlation of decay particles has been proposed \cite{Kienle:06}.  

The conventional methods to produce kaonic bound states are to use strangeness-transfer reactions of $(K^-,\pi^-)$, $(\pi^+,K^+)$, $(K^-,N)$ and $(\gamma,K^+)$. The formation of $\bar{K}$ clusters can be treated by a $\Lambda^*$ doorway model \cite{Yamazaki:02}, in which a $\Lambda^* (\equiv \Lambda_{1405})$ produced in elementary processes, typically,  
\begin{eqnarray}
K^- + n \rightarrow \Lambda^* + \pi^-, \label{eq:Kpi}\\
\pi^+ + n \rightarrow \Lambda^* + K^+,\label{eq:piK}
\end{eqnarray}
merges with a surrounding nucleon (or nucleus) to become a $\bar{K}$ state. The spectral shapes for the $d(\pi^+,K^+)K^-pp$ and $^3{\rm He}(\pi^+,K^+)K^-ppp$, calculated by this $\Lambda^*$ doorway treatment as given in Ref.~\cite{Akaishi:05}, are characterized by a large quasi-free component and a small bound-state peak. Typically, the bound-state formation fraction is of the order of 1 \%. 

In the present paper we study the possibility to make use of  the elementary process,
\begin{equation}
p + N \rightarrow p + \Lambda^* + K,
\end{equation}\label{eq:pp}
in connection with an experiment proposal at GSI using the FOPI detector \cite{FOPI-proposal}. Since the momentum transfer in this associated production of $\Lambda^*$ is very large ($Q \sim$ 1.6 GeV/$c$), one would expect that the formation cross section of $\bar{K}$ clusters must be very small. This process resembles the hypernuclear production process, $^A [Z](p, K^+)^{A+1}_{\Lambda}[Z]$, on a nuclear target, the cross section of which was evaluated by Shinmura {\it et al.}  \cite{Shinmura} to be $10^{-6}$ of the elementary $\Lambda$ production cross section, and this can be enhanced up to $10^{-4}$ when a short-range correlation is taken into account. On the other hand,  with a naive coalescence mechanism one obtains a sticking probability of the order of 0.1-1.0 \% because the internal momentum of the $\bar{K}$ clusters is very large \cite{Suzuki-Fabietti:05}. Still, most of primarily produced $\Lambda^*$ are expected to escape, and the quasi-free process dominates. We have studied this proton-induced associated production process more realistically, and found a surprisingly large production cross section by a unique mechanism, as described below.

\begin{figure*}[htb]
\centering
\includegraphics[width=16cm]{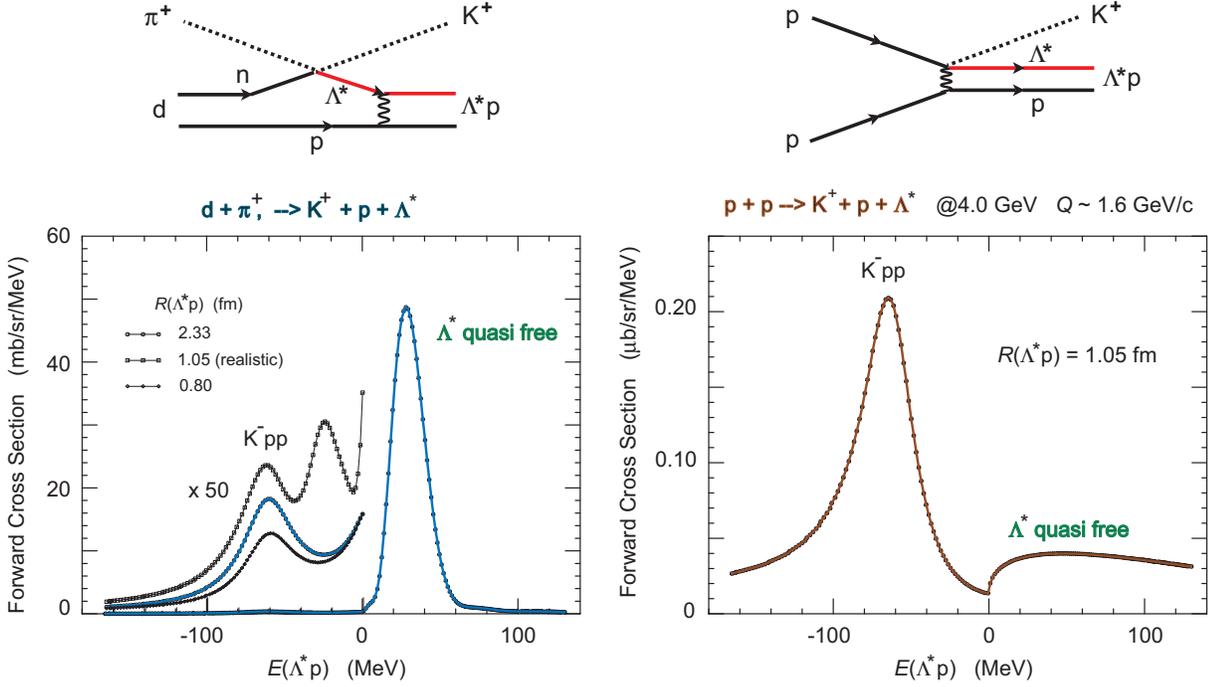}
\vspace{-0.5cm}
\caption{\label{fig:piK-pp-combined} 
(Upper) Diagram for $d(\pi^+,K^+)K^-pp$ and $p(p,K^+)K^-pp$. (Lower) Calculated spectral shapes of the $d(\pi^+,K^+)K^-pp$ and the $p(p,K^+)K^-pp$ reactions.
}
\end{figure*}

\section{Ordinary $\bar{K}$ transfer reactions}

We first treat conventional $\bar{K}$ transfer reactions, typically, $d(\pi^+,K^+)K^-pp$, based on the $\Lambda^*$ doorway model \cite{Yamazaki:02}. Hereafter, we use revised theoretical values for the  binding energy and width for $K^-pp$ ($B_K = 86$ MeV and $\Gamma = 58$ MeV). In the elementary process, Eq.(\ref{eq:piK}), the produced $\Lambda^*$ interacts with a proton in the target $d$, proceeding to $K^-pp$, as shown in Fig.~\ref{fig:piK-pp-combined} (Upper-Left). The momentum transfer at a typical incident momentum of $p_{\pi} \sim 1.5$ GeV/$c$ is $Q \sim$ 600 MeV/$c$. The energy spectrum involving both the bound and unbound regions was calculated following the Morimatsu-Yazaki procedure \cite{Morimatsu-Yazaki}. It is  given by 
\begin{eqnarray}
  \frac{{\rm d}^2\sigma}{{\rm d}E_{K^+} {\rm d} \Omega_{K^*}} &=& \alpha (k_{K^+}) \frac{{\rm d}\sigma ^{\rm elem}_{\Lambda^*}}{ {\rm d} \Omega_{K^+}} \nonumber \\
&\times&  \frac{| \langle \Lambda^*| V_{\bar{K}N}^{I=0}|\Lambda^* \rangle |^2}{\tilde{E}^2 + \frac{1}{4} \Gamma_{\Lambda^*}^2} S(E)
\end{eqnarray}\label{eq:piK-cross-section}
with a spectral function
\begin{eqnarray}
S(E)&=&(-\frac{1}{\pi}) {\rm Im} \big[\int {\rm d}\vec{r}_{K} {\rm d}\vec{r}'_{K} \tilde{f}^*(\vec{r}_{K}) \nonumber \\
&\times& \langle \vec{r}_{K} |\frac{1}{E-H_{{K}^-{\rm pp}} + i \epsilon}|\vec{r}'_{K} \rangle \tilde{f}(\vec{r}'_{K}) \big],
\label{eq:spectral}
\end{eqnarray} 
where
$\tilde{E}$ is the energy transfer to the $\Lambda^*$-$p$ relative motion in doorway states, 
 and   
$E$ the energy transfer to the $K^-$-$pp$ relative (internal) motion in the $K^-pp$ system, and $\alpha (k_{K^+})$ is a kinematical factor. The function  $\tilde{f}(r)$ is
\begin{equation}
 \tilde{f}(\vec{r}) = 2^3 {\rm e}^{{\rm i}2 \beta \vec{q}\vec{r}} C(r) \Phi_{pp}^* (2r) \Psi_{d} (2r)/|\Phi_{\Lambda^*} (0)|,
\end{equation}
with 
$\vec{q} = \vec{k}_{\pi^+} -\vec{k}_{K^+}$,   
$\beta = M_{p}/(M_{\Lambda^*} + M_{p})$ and $C(r) = 1 - {\rm exp}[-(r/1.2~{\rm fm})^2]$. In this derivation we have used a zero-range approximation for $V_{\bar{K} N} ^{I=0}$ and closure approximation to doorway states. 

The calculated spectral function is shown in Fig.~\ref{fig:piK-pp-combined} (Lower-Left). The dominant part is the quasi-free component, in which the produced $\Lambda^*$ escapes, and only a small fraction constitutes a bound-state peak. The figure shows that the bound-state peak intensity, though small, depends on the size of the $\Lambda^* p$ system.

\section{$\Lambda^* p$ doorway process in $NN$ collisions}

Now we consider the following process with a projectile proton and a target proton,
\begin{eqnarray}
p + p &\rightarrow& K^+ + \Lambda^* + p, \nonumber \\ 
         &\rightarrow& K^+ + \Lambda^* p, \nonumber \\
         &\rightarrow&  K^+ + K^-pp. \label{eq:pp2KLp}
\end{eqnarray}
where $K^- pp$ denotes a $\bar{K}$ cluster with structure of $K^- (pp)_{I=1,S=0}$ and decays subsequently as
\begin{eqnarray}
K^-pp &\rightarrow& \Lambda + p \rightarrow p + \pi^- + p, \label{eq:pp2Kpp-decay-1}\\
K^-pp &\rightarrow& \Sigma^0 + p \rightarrow p + \pi^- + \gamma + p, \label{eq:pp2Kpp-decay-2}\\
K^-pp &\rightarrow& \Sigma^+ + n \rightarrow n + \pi^+  + n. \label{eq:pp2Kpp-decay-3}
\end{eqnarray}
Its diagram is shown in Fig.~\ref{fig:piK-pp-combined} (Upper-Right). When the incident proton interacts with a neutron in a deuteron target, an analogous process takes place, namely, $p + n \rightarrow K^+ + \Lambda^*  + n$ with an iso-doublet partner $\Lambda^* n$ (=$K^- (pn)_{I=1,S=0}$) to be formed as well as $p + n \rightarrow K^0 + \Lambda^* + p$.  Hereafter, we take the $p + p$ case without loss of generality.

\begin{figure}[htb]
\centering
\vspace{-0.5cm}
\includegraphics[width=7.5cm]{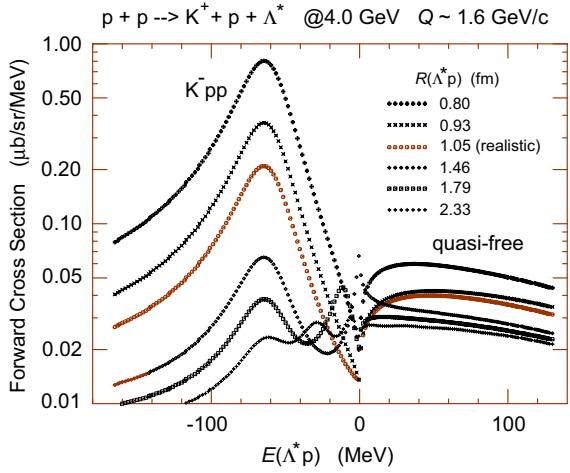}
\vspace{-1cm}
\caption{\label{fig:S(E)-parameters} 
Spectral shapes with different size parameters.}
\end{figure}

The $p \rightarrow p + K^- + K^+$ process, where a $K^+K^-$ pair is assumed to be created at zero range from a proton, is of highly off-energy shell ($\Delta E \sim 2 m_K$). This process is realized with a large momentum transfer to the second proton, which occurs efficiently by a short-range $pp$ interaction, expressed by a Yukawa type interaction exp$(-m_B/r)/r$ with $m_B$ being the intermediate boson mass ($m_B$). Then, the effective interaction for the elementary process is written as 
\begin{eqnarray}
&& \langle\vec{r}_{K^+(K^- pp')}, \vec{r}_{(K^- p)p'}, \vec{r}_{K^-p}| t | \vec{r}_{pp'} \rangle \nonumber\\
&&= T_0 \int {\rm d} \vec{r} \, F(\vec{r}) \delta (\vec{r}_{K^+(K^- pp')} - \eta \vec{r}) \nonumber\\
&& \times~ \delta(\vec{r}_{(K^- p)p'} - \vec{r}) 
 \delta(\vec{r}_{K^- p}) \delta(\vec{r}_{p p'} - \vec{r})),
\end{eqnarray}
where
\begin{equation}
F(\vec{r}) = \frac{\beta}{r} \, {\rm exp}(-\frac{r}{\beta})
\end{equation}
with $\beta = \hbar/(m_B c) $ and $\eta = M_p/M_{K^-pp}$. 
The $\Lambda^*$ is treated as a quasi-bound state of $K^- p$, and the interaction matrix element for $\Lambda^*$ formation is given by 
\begin{eqnarray}
&& \langle \vec{k}_{K^+ (\Lambda^* p')} , \vec{r} = \vec{r}_{\Lambda^* p'}, \phi_{\Lambda^*} | t | \vec{k}_{p p'} \rangle \nonumber\\
&&
= T_0 \, \phi_{\Lambda^*} (0)\, F(r)  \langle \vec{k}_{K^+(\Lambda^* p')} | \eta \vec{r} \rangle \langle\vec{r} | \vec{k}_{pp'}\rangle \nonumber\\
&& \equiv U_0 f (\vec{r}),
\end{eqnarray}
where 
\begin{eqnarray}
U_0 &=& \frac{1}{(2 \pi)^3} T_0 \phi_{\Lambda^*} (0),\\
f(\vec{r}) &=& \frac{\beta}{r} {\rm exp} (-\frac{r}{\beta} + {\rm i} \vec{Q} \vec{r}),\\
\vec{Q} &=& \eta_0 \vec{k}_p + \eta \vec{k}_{K^+}, \\
\eta_0 &=& \frac{1}{2} + \frac{m_K}{M_{K^-pp} + m_K} \eta.
\end{eqnarray}
The production cross section of $\Lambda^*p$ ($= K^- pp$) is given by
\begin{eqnarray}
&&\frac{{\rm d}^3 \sigma} {{\rm d} E_{K^+}  {\rm d} \Omega_{K^+}} = \frac{(2\pi)^4}{(\hbar c)^2} |U_0|^2 \frac{k_{K^+} E_p}{2\, k_p} (-\frac{1}{\pi})  \nonumber\\
&& \times~ {\rm Im} \big[\int \! \int {\rm d} \vec{r}' {\rm d} \vec{r} \nonumber \\
&& \times~ f^*(\vec{r}' ) \langle \vec{r}'|\frac{1}{E - H_{\Lambda^*p}}|\vec{r} \rangle f(\vec{r}) \big]
\end{eqnarray}
where
\begin{eqnarray}
&&E = E_p - E_{K^+} - M_{\Lambda^*} c^2 - E^{\rm Recoil}_{K^- pp}, \nonumber \\
&&H_{\Lambda^* p} = -\frac{\hbar^2}{2 \, \mu_{\Lambda^* p}} \vec{\nabla}^2 + (V_0 + {\rm i} W_0) {\rm exp} (-\frac{r^2}{b^2}) \nonumber
\end{eqnarray}
with
$V_0 = - 295$ MeV, $W_0 = -42$ MeV, and $b = 1.0$ fm.
Then, the spectral function of $\Lambda^*p$ ($= K^- pp$) is
\begin{eqnarray}
&&S(E) = - \frac{1}{\pi}  {\rm Im} \big[\int \! \int {\rm d} \vec{r}' {\rm d} \vec{r} f^*(\vec{r}') \nonumber \\
&& \times \langle \vec{r}'|\frac{1}{E - H_{\Lambda^*p}}|\vec{r}\rangle f(\vec{r}) \big] \nonumber\\
&& = -\frac{8 \mu_{\Lambda^* p}}{\hbar^2} \sum_{l=0}^{\infty} (2 l +1) {\rm Im}  \big[\frac{1}{W(u_l^{(0)} u_l^{(+)})} \nonumber\\
&& \times \int_{0}^{\infty} {\rm d}r' \int_{0}^{\infty} {\rm d}r\, {\rm exp}(-\frac{r'}{\beta}) \, j_l (Q r') \nonumber\\
&& \times~  u_l^{(0)} (r_<) u_l^{(+)} (r_>) \, j_l (Qr) ~{\exp}(-\frac{r}{\beta}) \big],
\end{eqnarray}
where $u_l^{(0)}$ and  $u_l^{(+)}$ are the stationary and outgoing solutions of the Schr\"odinger equation,
and $W$ is the Wronskian of them.

  Essentially, the spectral function is approximately composed of the following three factors:
\begin{equation}
 T(r) \propto \frac{{\rm e}^{-m_B\, r}}{r} \times {\rm e}^{i \vec{Q} \vec{r}} \times G (r),
\end{equation}
where
\begin{equation}
G (r) = \big[ -{\rm Im} \{\frac{u^{(0)} (r)\, u^{(+)} (r)}{W(u^{(0)} \, u^{(+)} )}  \} \big]^{1/2}.
\end{equation}
They are: i) the collision range $1/m_B$, ii) the momentum transfer $Q$ and iii) the structure function $G(r)$ depending on the rms distance $R(\Lambda^* p)$ of the $\Lambda^*-p$ system. The calculated wavefunction of $K^- pp$  yields $R(\Lambda^* p) =$ 1.05 fm. The momentum transfer in the reaction is $Q$ = 1.6 GeV/$c$. The boson mass in producing $\Lambda^*$ in $pp$ collision is taken to be the $\rho$ meson mass; $m_B =  m_{\rho} =  770$ MeV/$c^2$.

The calculated spectral functions (forward cross sections) at $T_p$ = 4 GeV are presented in the scale of $E(\Lambda^* p) = 27~{\rm MeV} - B_K$ in Fig.~\ref{fig:piK-pp-combined} (Lower-Right). The cross section integrated over the quasi-free region ($E(\Lambda^* p) > 0$) corresponds to the free emission of $\Lambda^*$, which is known to have an empirical cross section of  $\sigma (pp \rightarrow K^+ + \Lambda^* + p) =$ 20 $\mu$b from a DISTO experiment at $T_p$ = 2.85 GeV \cite{DISTO}.  So, we have adjusted our absolute cross sections so as to give this empirical cross section. 

Surprisingly, in great contrast to the ordinary cases, the spectral function is peaked at the bound state with a very small quasi-free component. This dominance of $\Lambda^* p$ sticking in such a large-$Q$ reaction can be understood as originating from the matching of the small impact parameter with the small size of the bound state. For further understanding of the mechanism we examined the dependence of the spectral function by changing the essential parameters fictitiously. Fig.~\ref{fig:S(E)-parameters} (Upper) shows that the bound-state peak decreases dramatically, when we increase the rms size $R(\Lambda^* p)$ from 1.05 fm (the predicted size of the dense $ppK^-$) to 1.46 and 2.33 fm.  It also shows that with a denser system ($R(\Lambda^* p)$ = 0.93 and 0.80 fm) the peak height increases. So, the dominant sticking of $\Lambda^* p$ is the result of the dense $\bar{K}$ system to be formed. 

\begin{figure*}
\centering
\vspace{0cm}
\includegraphics[width=15cm]{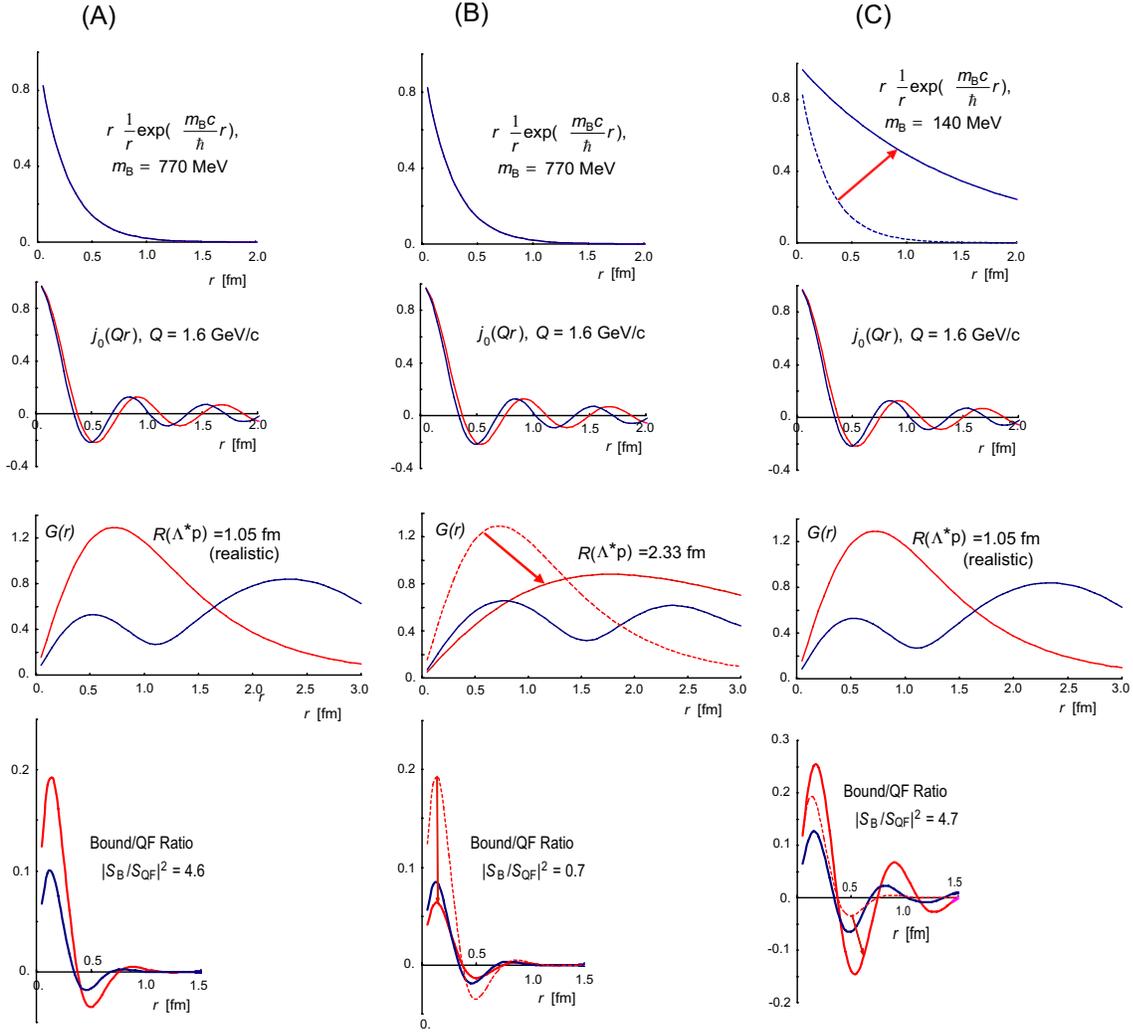}
\vspace{0cm}
\caption{\label{fig:T-amplitude} 
Effects of various parameters on the transition amplitude in the $p + p \rightarrow K^+ + K^-pp$. The solid and broken curves represent the bound ($E_{\rm B}(\Lambda^*p) = -62$ MeV) and the quasi-free ($E_{\rm QF} (\Lambda^*p)$ = 100 MeV) cases. }
\end{figure*}

We also showed that with a fictitiously long-range $NN$ collision ($m_B$ = 10 MeV) the bound-state peak diminishes and the quasi-free component dominates. Under this condition the bound-state peak is enhanced only when the momentum transfer is small (so called recoilless condition). On the other hand, the dominant sticking of $\Lambda^* p$ is assisted by the large momentum transfer ($Q \sim $ 1.6 GeV/$c$). 

As a more analytical way to show the physics behind we plot the radial dependence of each factor of the transition intensity in Fig.~\ref{fig:T-amplitude}. The first row shows the $p-p$ interaction range of the Yukawa type with different boson mass $m_B$. The second row shows the spherical Bessel function $j_0 (Qr)$ corresponding to the momentum transfer of $Q =$ 1.6 GeV/$c$. These two functions have values mostly for $r < $ 0.3 fm. The third row shows the structure dependent function $G(r)$, where the solid and broken curves represent the bound ($E_{\rm B}(\Lambda^*p) =$ -62 MeV) and the quasi-free ($E_{\rm QF} (\Lambda^*p)$ = 100 MeV)  cases, respectively. Finally, the bottom row shows the radial dependences of the spectral strengths at the bound-state (B) and the quasi-free (QF) regions. In the realistic case,  Column (A), the structure function is dense below $R(\Lambda^* p)$ = 1.05 fm so that it overlaps with the short range interaction assisted by the large momentum transfer. The Bound/QF ratio is 4.6. If we  artificially increase the $\Lambda^*p$ distance to 2.33 fm (a deuteron like distance), as shown in Column (B) the overlap drops down so that the ratio becomes 0.7. As far as the momentum transfer is large, a softened interaction ($m_B =$ 140 MeV), Case (C), does not change the Bound/QF ratio from the realistic case (A).    

We have thus demonstrated that the dominant sticking of $\Lambda^* p$ occurs as a joint effect of the short-range collision, the large momentum transfer and the compact size of the $\bar{K}$  cluster. It is vitally important to examine our results experimentally. An experimental observation of $K^- pp$ in $pp$ collision will not only confirm the existence of $K^- pp$, but also proves the compactness of the $\bar{K}$ cluster.  
  
\section{Subsequent processes}

Once a $\Lambda^* p$ doorway is formed, the $\Lambda^*$ cannot escape any more as a free particle and thus it is likely to further propagate in a complex nucleus as
\begin{eqnarray}
&&\Lambda^* p + ``p" \rightarrow K^-ppp,\\
&&\Lambda^* p + ``n" \rightarrow K^-ppn,
\end{eqnarray}
and ultimately a {\it kaonic proton capture} reaction may occur, such as
\begin{eqnarray}
&&d(p,K^+)K^-ppn,\\
&&d(p,K^0)K^-ppp,\\
&&^3 {\rm He}(p,K^+)K^-pppn,\\
&&^3 {\rm He}(p,K^0)K^-pppp.
\end{eqnarray}
In principle, missing mass spectra, $MM(K^+)$ and $MM(K^0)$, may reveal monoenergetic peaks. 

\section{$p+p \rightarrow K^+ + p + \Lambda$ reaction} 

The elementary reaction of type 
\begin{equation}
p + p \rightarrow K^+ + Y^0 + p
\end{equation}
was studied experimentally by the DISTO group at SATURNE \cite{DISTO} and more recently by the ANKE group at COSY \cite{Zychor:06}. The DISTO experiment identified $\Lambda$ from  the invariant-mass spectrum of $p + \pi^-$ and constructed a missing mass spectrum of $K^+ p$ from those events involving $\Lambda$ at an incident proton energy of $T_p$ = 2.85 GeV. The missing mass in this case is equal to the mass of $Y^0$, 
\begin{equation}
MM(K^+p) = M(Y^0),
\end{equation}
and they found peaks associated with the production of $\Lambda(1115)$, $\Sigma^0 (1193)$ and ${\Sigma^0} (1385) + \Lambda (1405)$. They obtained a cross section of 20 $\mu$b for $\Lambda (1405)$, which we used as an absolute scale in our calculations.

The ANKE experiment measured the energies and momenta of four emitted particles, $K^+$, $p$, $X^{+,-}$ and $\pi^{-,+}$, with complete kinematical constraint at $T_p$ = 2.83 GeV:
\begin{equation}
p + p \rightarrow K^+ + p + Y^0 \rightarrow K^+ + p + X^{+,-} + \pi^{-,+}.
\end{equation}
They constructed $MM(K^+ p)$ and additionally $MM(K^+ p \pi^-)$, which was equated to $M(X)$. When $MM(K^+ p \pi^-) = M(p)$, it indicates that this $Y^0$ decays to $p + \pi^-$, and thus, it is assigned to be $\Lambda$. Thus, the ANKE $MM(K^+ p)$ spectrum is equivalent to the DISTO spectrum. 

So far, the reaction products were studied in terms of the elementary processes. Now, we propose to examine the new situation related to the existence of $K^- pp$. We point out that the formation/decay process of this object, as given in eq.(\ref{eq:pp2KLp}) and (\ref{eq:pp2Kpp-decay-1},\ref{eq:pp2Kpp-decay-2},\ref{eq:pp2Kpp-decay-3}), are hidden in the observed spectra of $MM(K^+ p)$. The most important information is contained in a spectrum of $MM(K^+)$, which is related to the mass of $K^- pp$,
\begin{equation}
MM(K^+) = M(K^- pp),
\end{equation}      
but no such spectrum has been obtained. Presumably, the limited detector acceptance may not allow to reach the mass range of the present concern ($M(K^-  pp) \sim 2280$ MeV/$c^2$). 
Now, a new experiment of the FOPI group at GSI, which is aimed at measuring the whole products in the $p + p$ reaction at $T_p$ = 3.5 GeV to reconstruct both the invariant mass $M_{inv} (\Lambda p)$ and $MM(K^+)$, is in progress. The FOPI group is also studying the $p + d$ process, which may lead to the formation of $K^-ppn$ cluster.  

\section{Concluding remarks}

In this note we have shown that the basic $\bar{K}$ cluster, $K^-pp$, as a dissolved state of $\Lambda^* + p$ can be formed with extraordinarily large enhancement, since the $\Lambda^*$ produced in a short-range collision with the participating proton automatically forms a $\Lambda^* p$ doorway toward a dense $K^-pp$ system. This anomalous dominance of $\Lambda^* p$ sticking results from the unusual matching of the short  collision range ($1/m_B \sim$ 0.3 fm) and the small radius of the produced $K^- pp$ ($R(\Lambda^* p) \sim$ 1.0 fm), assisted by a large momentum transfer. Experimental confirmation of this effect will automatically proves the dense character of $\bar{K}$ clusters. 

Finally, we comment on our earlier proposal to make use of hot fireball in heavy-ion collisions as sources of various $\bar{K}$ clusters \cite{Yamazaki:04}. If the formation of $\Lambda^* p$ doorway toward $K^-pp$ is really enhanced in the $NN$ collisions, the $\bar{K}$ cluster formation in heavy-ion reactions is expected to be enhanced as well, since a heavy-ion reaction involves lots of primary $NN$ collisions and subsequent processes.  

We would like to thank Prof. S. Shinmura for the illuminating discussion, and Prof. P. Kienle, Prof. N. Herrmann, Dr. K. Suzuki and Dr. L. Fabietti for the daily stimulating discussion during the course of the experimental collaboration.  We acknowledge the receipt of Grant-in-Aid for Scientific Research of Monbu-Kagakusho of Japan.

\end{document}